\documentclass{article}
\usepackage{spconf,amsmath,graphicx}
\usepackage[htt]{hyphenat}
\usepackage{booktabs}
\usepackage{makecell}
\usepackage{enumitem}
\usepackage{multirow}
\usepackage{float}
\usepackage[labelformat=simple]{subcaption}

\usepackage[font=small,labelfont=bf]{caption}
\usepackage{flushend}
\usepackage{tikz}
\usepackage{hyperref}

\title{PROBING THE INFORMATION ENCODED IN X-VECTORS}

\name{Desh Raj, David Snyder, Daniel Povey, Sanjeev Khudanpur\thanks{This work was partially supported by NSF CRI Grant No 1513128, and DARPA LORELEI Contract No HR0011-15-2-0024.}}
\address{Center for Language and Speech Processing \& Human Language Technology Center of Excellence\\The Johns Hopkins University, Baltimore, MD 21218, USA.\\\footnotesize{\texttt{draj@cs.jhu.edu, \string{david.ryan.snyder, dpovey\string}@gmail.com, khudanpur@jhu.edu}}}

\begin{document}
%
\maketitle
\begin{abstract}
Deep neural network based speaker embeddings, such as x-vectors, have been shown to perform well in text-independent speaker recognition/verification tasks. In this paper, we use simple classifiers to investigate the contents encoded by x-vector embeddings. We probe these embeddings for information related to the speaker, channel, transcription (sentence, words, phones), and meta information about the utterance (duration and augmentation type), and compare these with the information encoded by i-vectors across a varying number of dimensions. We also study the effect of data augmentation during extractor training on the information captured by x-vectors. Experiments on the RedDots data set show that x-vectors capture spoken content and channel-related information, while performing well on speaker verification tasks.
\end{abstract}
\section{Introduction}
\label{sec:intro}

In the last few years, speaker embeddings, especially those extracted from deep neural networks (DNNs) discriminatively trained to classify speakers, have become very popular for tasks like speaker identification and verification. In several studies, DNNs have been used to replace the universal background models (UBMs)~\cite{lei2014novel}, to obtain bottleneck features in addition to traditional features~\cite{mclaren2015advances}, and for text-dependent~\cite{variani2014deep} and text-independent~\cite{snyder2016deep} speaker verification.

In particular, x-vectors~\cite{snyder2017deep} have been shown to obtain state-of-the-art performance on text-independent speaker verification. In this paper, we investigate whether an x-vector embedding, which is trained solely to predict the speaker label, also contains incidental information about the transcription, channel, or meta-information about the utterance. Toward this objective, we design simple classification and regression based probing tasks which examine the embedding for these properties. It was shown in~\cite{snyder2018x} that x-vector systems effectively exploit data augmentation strategies for improved performance on speaker recognition. Our investigation using x-vector extractors trained with and without augmentation sheds some light on the possible reason behind this improvement. Previous work has shown that i-vectors, though developed for speaker recognition, can improve automatic speech recognition (ASR), because they capture speaker and channel characteristics \cite{saon2013speaker}. Our probing task results suggest that x-vectors also capture similar information and hence motivate their use for speaker adaptation in ASR.

Wang et al.~\cite{wang2017does} have previously conducted similar investigations for i-vectors~\cite{dehak2010front} and d-vectors~\cite{variani2014deep}. However, their probing tasks were designed on the RSR2015 data set~\cite{larcher2012rsr2015}, while we use the RedDots data set\footnote{RedDots is freely available for academic research from \url{https://sites.google.com/site/thereddotsproject/home}.}~\cite{lee2015reddots}. More recently, it was shown that both i-vectors and x-vectors contain information about the speaking style and emotion~\cite{williams2019disentangling}. In natural language processing (NLP), probing tasks for embeddings have gained attention~\cite{conneau2018you} due to sentence encoders such as BERT~\cite{devlin2018bert}, which are pretrained on language modeling, but achieve state-of-the-art performance across several other tasks.

Since x-vectors are trained in a text-agnostic fashion by predicting the speaker label given the input utterance features, they perform well in text-independent speaker verification tasks. However, they may also be incorporated into a text-dependent system, if used in conjunction with a keyword spotting component. We demonstrate this by presenting results for text-dependent speaker verification using x-vectors trained in a text-independent manner and compare this method with an i-vector based system. These experiments are conducted on the RedDots data set.

The remainder of this paper is organized thus. We first describe the theory and our implementation of i-vector and x-vector speaker embeddings in Section~\ref{sec:embeddings}, followed by a description of the RedDots data set, the probing tasks, and classifiers for the probing tasks in Section~\ref{sec:task}. In Section~\ref{sec:experiment}, we analyze the results of the probing tasks for i-vector and x-vector embeddings of different dimensions. Finaly, we present results for our x-vector based system on text-dependent speaker verification in Section~\ref{sec:reddots}, and conclude in Section \ref{sec:conclusion}.  

\section{Speaker embeddings}
\label{sec:embeddings}

In this section, we describe the speaker embeddings, i.e., i-vectors and x-vectors. Both these systems were built using the Kaldi speech recognition toolkit~\cite{povey2011kaldi}, and trained on the \textit{dev} portion of \textit{VoxCeleb2}~\cite{Chung18voxceleb}, which consists of 1,092,009 utterances from 5,994 distinct speakers.

\subsection{i-vector}

First proposed in \cite{dehak2010front}, the i-vector framework assumes that the speaker and session-dependent supervector $\mathbf{M}$ of Gaussian mean vectors may be modeled as
\begin{equation}
\mathbf{M} = \mathbf{m} + \mathbf{Tw},
\end{equation}
where $\mathbf{m}$ is the speaker and session-independent supervector obtained from a Gaussian mixture model (GMM) based universal background model (UBM), $\mathbf{T}$ is a low-rank total variability matrix that captures both speaker and session variability, and the i-vector is the posterior mean of $\mathbf{w}$. 

Our implementation of the traditional GMM-UBM based i-vector system is similar to that described in~\cite{snyder2015time}. The system is trained on 30 MFCC features with a frame-length of 25ms that are mean-normalized over a sliding window of up to 3 seconds. An energy-based speech activity detection (SAD) system selects features corresponding to speech frames. The UBM is a 2048 component full-covariance GMM.

\subsection{x-vector}

Our x-vector system is similar to that implemented in \cite{snyder2017deep}, and described in detail there\footnote{An example recipe is in the main branch of Kaldi at \texttt{https:
//github.com/kaldi-asr/kaldi/tree/master/egs/
voxceleb/v2}}. The features are 30 dimensional filterbanks with a frame-length of 25ms, mean-normalized over a sliding window of up to 3 seconds. The same energy SAD as used in the i-vector system filters out nonspeech frames.

\begin{table}[tb]
\centering
\begin{tabular}{cccc}
\toprule
\thead{\textbf{Layer}} & \thead{\textbf{Layer context}} & \thead{\textbf{Total}\\\textbf{context}} & \thead{\textbf{Input $\times$ Output}} \\
\midrule
frame1 & [$t-2$, $t+2$] & 5 & $150\times 512$ \\
frame2 & \{$t-2$, $t+2$\} & 9 & $1536\times 512$ \\
frame3 & \{$t-3$, $t+3$\} & 15 & $1536\times 512$ \\
frame4 & $t$ & 15 & $512\times 512$ \\
frame5 & $t$ & 15 & $512\times 512$ \\
stats\_pooling & [$0,T$) & $T$ & $1500T\times 3000$ \\
segment6 & \{0\} & $T$ & $3000\times d$ \\
segment7 & \{0\} & $T$ & $d\times 512$ \\
softmax & \{0\} & $T$ & $512\times N$ \\
\bottomrule
\end{tabular}
\caption{The DNN architecture for the x-vector system. Embeddings are extracted at layer \textit{segment6}, before the nonlinearity. Here, $T$, $d$, and $N$ denote the number of utterance frames, the dimensionality of embedding required, and the number of speakers, respectively.}
\label{tab:xvec_dnn}
\end{table}

Table~\ref{tab:xvec_dnn} describes the DNN architecture for the x-vector system. The first 5 layers, called \textit{frame} layers, operate at the context-dependent frame level. For example, the input for \textit{frame2} is the spliced output of \textit{frame1} at frames $t-2$, $t$, and $t+2$. Since the context for \textit{frame1} extended from $t-2$ to $t+2$, this means that the total context for \textit{frame2} is 9.

At the statistical pooling layer, \textit{frame5} outputs from all $T$ frames are aggregated by computing the mean and standard deviation. The subsequent frames operate on this 1500-dimensional vector which represents the entire segment, and are named \textit{segment} layers. Since we extract embeddings at \textit{segment6}, the output dimension at this layer is set to the required dimension, say $d$. The final softmax layer has $N$ output nodes, where $N$ is the number of speakers in the training data. All nonlinearities are rectified linear units (ReLUs).
 
\section{Probing tasks}
\label{sec:task}

The underlying assumption for using probing tasks is that if information pertaining to a property is encoded by the embedding, we can train a classifier to predict this property, and the performance of the classifier is proportional to the amount of information that the embedding encodes~\cite{wang2017does}. In our experiments, we focus on properties related to the speaker, the utterance transcription, and meta information about the utterance.

\subsection{Data set}
\label{sec:data}

We use the RedDots data set~\cite{lee2015reddots} with some modification for our probing tasks. Our primary reason for using this data is that it contains both common and unique utterances across a variety of speakers. We use the enrollment utterances corresponding to the common, unique, and free-choice pass-phrase portions of the data set. This gives us a total of 2484 utterances, comprising 460 unique transcriptions. We further apply augmentation on this data by adding music, speech, and noise using the MUSAN data set~\cite{snyder2015musan}, which consists of over 900 noises, 42 hours of music from various genres, and 60 hours of speech from twelve languages. This increases the total number of utterances to 9936. The final distribution of the frequency of unique utterances is given in Table~\ref{tab:utt_dist}.

\begin{table}[t]
\centering
\begin{tabular}{ccc}
\toprule
\textbf{\# unique utterances} & \textbf{Frequency} & \textbf{Total} \\
\midrule
8 & 456 & 3648 \\
2 & 444 & 888 \\
450 & 12 & 5400 \\
\bottomrule
\end{tabular}
\caption{Distribution of frequency of unique utterances. The utterances with high frequency correspond to the common pass-phrases. For the utterances with lower frequency, each of these are spoken by 3 different speakers, and 4-fold augmentation gives a total of 12.}
\label{tab:utt_dist}
\end{table}

\subsection{Probing Tasks}

We investigate the speaker embeddings using 8 different tasks that are designed to probe the speaker, transcription, and utterance-level meta information encoded in the embedding.

\begin{enumerate}[leftmargin=*]
\setlength\itemsep{0pt}
\item \textbf{Session identification}: This task probes the ability of the embedding to predict the session, which is a more fine-grained recognition task than speaker identification. Since different sessions for the same speaker may have different acoustic characteristics owing to channel effects or varying background conditions, we conjecture that embeddings which predict the session well may encode some channel information. The data set comprises 828 different sessions in total.
\item \textbf{Speaker gender}: This task investigates whether the embedding can distinguish between genders, i.e., a binary classification task. Although it is inherently an easier task than speaker identification, our data is imbalanced in the gender labels (49 male and 13 female speakers).
\item \textbf{Speaking rate}: We augment all 9936 utterances by 3-way speed perturbation with rates 0.5, 1.0, and 1.5. A multi-class classifier with 3 classes is then trained on 80\% of this data and evaluated on the remaining 20\%. We report the overall accuracy to predict whether the speaker embeddings can encode information about the speaking rate.
\item \textbf{Transcription}: This task explores the ability of the speaker embedding to predict the exact transcription given the utterance. Although there are 460 different utterances, we select only the 100 most common ones for the prediction task. Still, there remains some imbalance in this set as well, since the top 10 most frequent utterances are the common phrases and occur much more frequently than the other utterances.
\item \textbf{Word recognition}: This task probes whether the embeddings capture information about words in the utterance. We select the top 50 most frequent words in the vocabulary and build a classifier for each word that predicts, given the embedding as input, whether the utterance contains the word. Finally, we compute, for each utterance, the fraction of words it labeled correctly, and then obtain the mean across all utterances.
\item \textbf{Phoneme recognition}: This task investigates whether the embeddings encode phone-level information in an utterance. For this, we select only the 24 consonant phonemes in the English language, since vowel phonemes are present in most utterances (regardless of length). Furthermore, since most such phonemes would still be present in most utterances, we only say that an utterance contains a phoneme if it contains at least $k=3$ occurrences of the phoneme~\footnote{This choice of $k$ is not arbitrary; it was tuned so as to get a sufficiently balanced training set.}. Similar to the word recognition task, we build a binary classifier for each such phoneme and finally compute the mean accuracy across all utterances. We use the \texttt{g2p}\footnote{\texttt{https://github.com/Kyubyong/g2p}} library for converting graphemes to phonemes.
\item \textbf{Utterance length}: In this task, we examine whether the x-vector embedding can predict the duration of the utterance. Since the duration is continuous, we model this as a regression task as opposed to the other probing tasks. The performance of the embedding is reported in terms of $1 - \frac{RMSE}{\sigma}$, where $RMSE$ is the root mean squared error, and $\sigma$ is the standard deviation of utterance lengths. This quantity can be interpreted as the percentage of variance explained by the embedding.
\item \textbf{Augmentation type}: Finally, this task investigates whether the type of augmentation (music, speech, noise, or no augmentation) is captured in the speaker embedding. We train a four-class classifier and report the accuracy of recognition.
\end{enumerate}

\begin{figure*}[t]
\begin{subfigure}{0.33\linewidth}
\centering
\includegraphics[width=\linewidth]{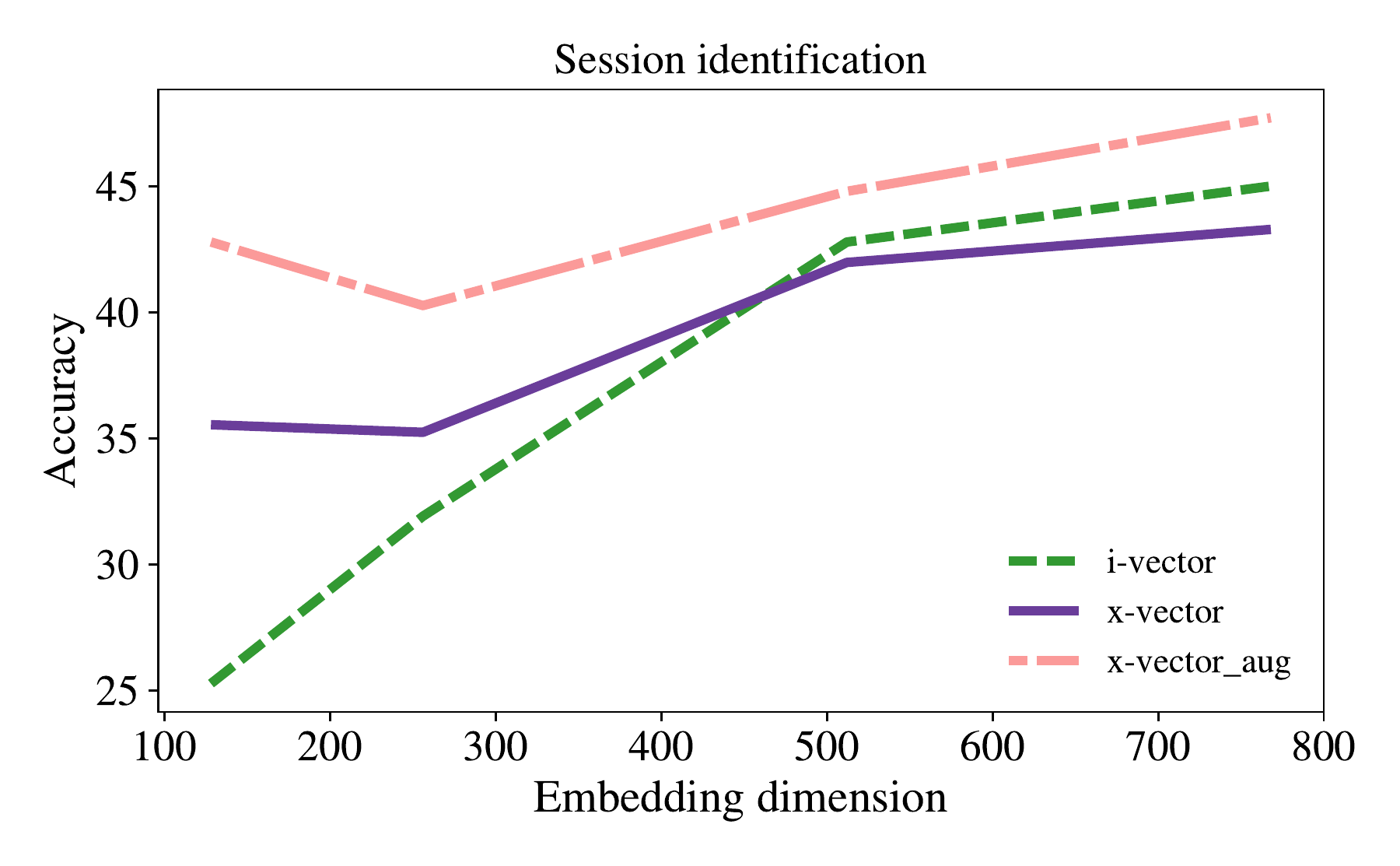}
\caption{}
\label{fig:sess_id}
\end{subfigure}
\begin{subfigure}{0.33\linewidth}
\centering
\includegraphics[width=\linewidth]{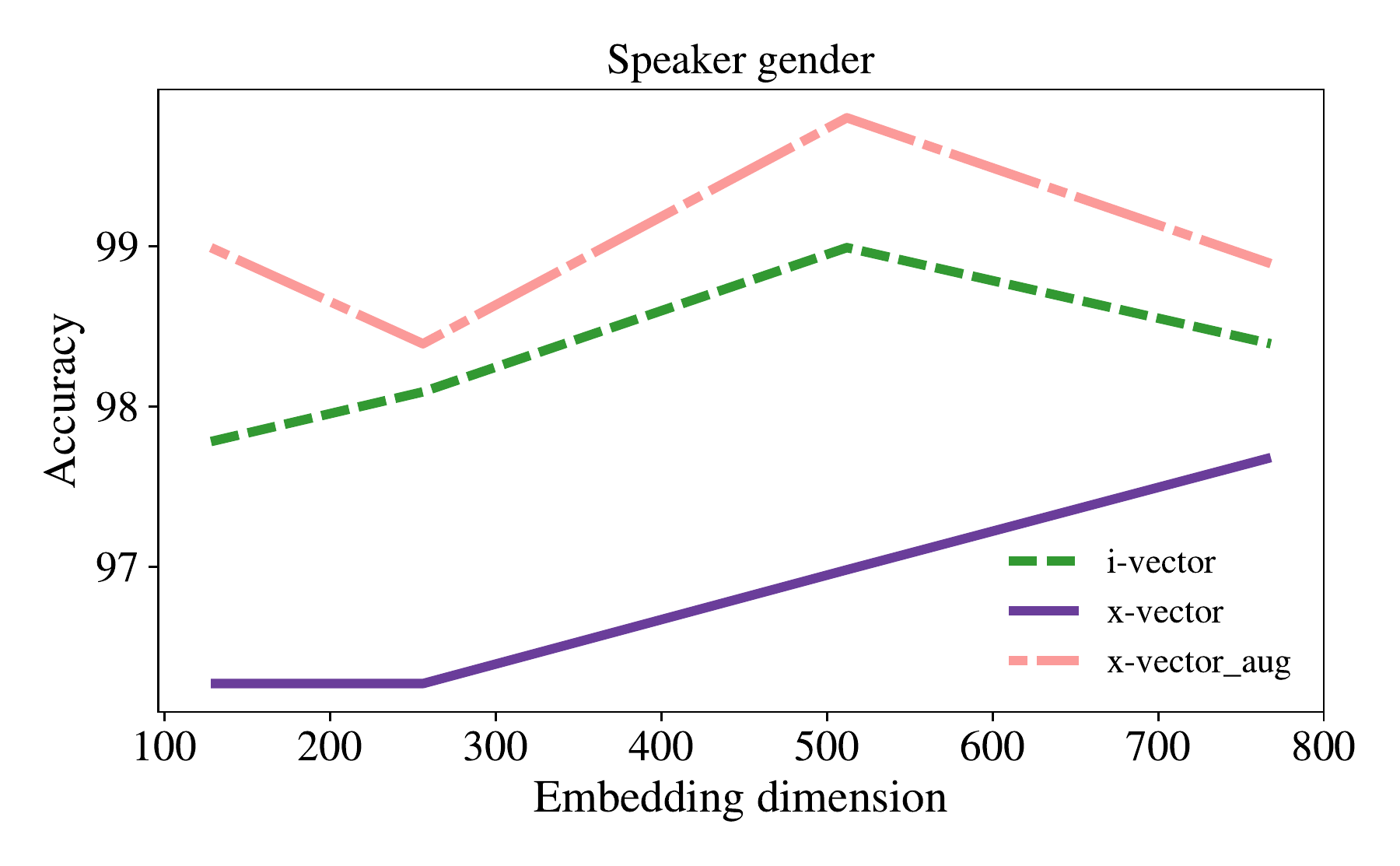}
\caption{}
\label{fig:gender}
\end{subfigure}
\begin{subfigure}{0.33\linewidth}
\centering
\includegraphics[width=\linewidth]{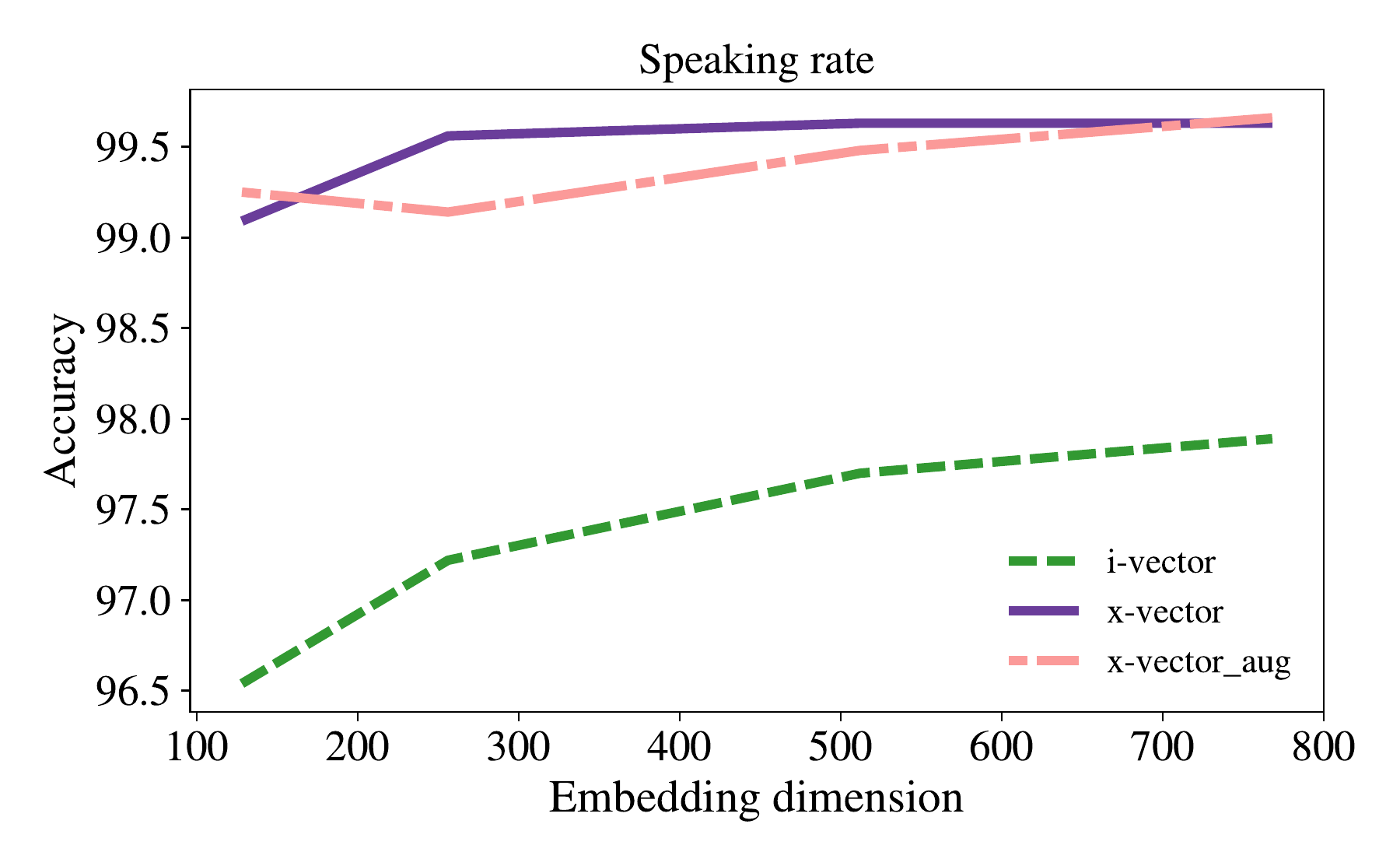}
\caption{}
\label{fig:speed}
\end{subfigure}\hfill
\caption{Performance of speaker embeddings in speaker related tasks: (a) session identification, (b) speaker gender, and (c) speaking rate recognition, across several dimensions.}
\label{fig:probe_res_spk}
\end{figure*}

\subsection{Classifiers}
\label{sec:classifier}

Since our objective is to evaluate the embeddings, we use a very simple classifier---a Multilayer perceptron (MLP) with a single hidden layer and ReLU activations. The hidden layer size is fixed at 500 for all the probing tasks. This system is implemented in Python using the PyTorch deep learning framework~\cite{paszke2017automatic}. We use a cross-entropy loss for the classifiers and a mean squared error loss for the regression task. Adam~\cite{kingma2014adam} is used as the optimizer in all probing tasks, with a learning rate of 0.001. Additionally, for the speaker gender task, we weigh the labels in inverse proportion to their sample sizes in the loss function, to account for the class imbalance. Unless stated otherwise, we train our classifiers on 90\% of the augmented RedDots data set and test on the remaining 10\%.

\section{Results and discussion}
\label{sec:experiment}

We now present the results of the probing tasks described in Section~\ref{sec:task}. For each task, we experiment with embeddings of dimensions 128, 256, 512, and 768. For each dimension, we have an i-vector and an x-vector extractor trained on unaugmented VoxCeleb2 data. Additionally, we investigate the effect of data augmentation on the information encoded by x-vectors. We apply an augmentation strategy based on~\cite{snyder2018x}, which doubles the amount of training data. We extend the clean VoxCeleb data with a second copy that has been randomly augmented with either noises from MUSAN~\cite{snyder2015musan} or convolved with simulated room impulse responses (RIRs)~\cite{ko2017study}. These are labeled as \texttt{x-vector\_aug} in the figures.

\subsection{Session identification}

In speaker recognition, it is desirable for embeddings to be invariant with channel or session characteristics. In contrast, capturing these sources of variability is beneficial for ASR, where acoustic models utilize embeddings (usually i-vectors) to adapt to both speaker and channel characteristics. Here we measure the amount of session information retained in the embeddings, by examining their performance on a session identification task (from 848 possible sessions). In Figure~\ref{fig:sess_id}, we see that while x-vectors significantly outperform i-vectors at low dimensions, they achieve a similar accuracy (∼45\%) at high dimensions. We observed that most of the errors were due to the embeddings attributing a session to another session from the same speaker, and analysis revealed that in these cases, the sessions had similar acoustic characteristics and transcriptions. We believe that this ability of x-vectors to capture speaker and channel characteristics makes them suitable for speaker adaptation in ASR systems.

\subsection{Speaker gender}

Both embeddings achieve high accuracy in recognizing gender. In Figure~\ref{fig:gender} we see that, without training data augmentation, i-vectors clearly outperform x-vectors in this task. On further analysis, we found that while the recall for female-labeled utterances was ∼ 1 for i-vectors, it was slightly lower for x-vectors, which leads to this difference in overall performance. However, once it is trained with augmentation, the ability of x-vectors to discriminate between speakers increases, and subsequently its gender identification performance reaches close to 99\% accuracy.

\begin{figure}[tb]
\begin{subfigure}{0.49\linewidth}
\centering
\includegraphics[width=\linewidth]{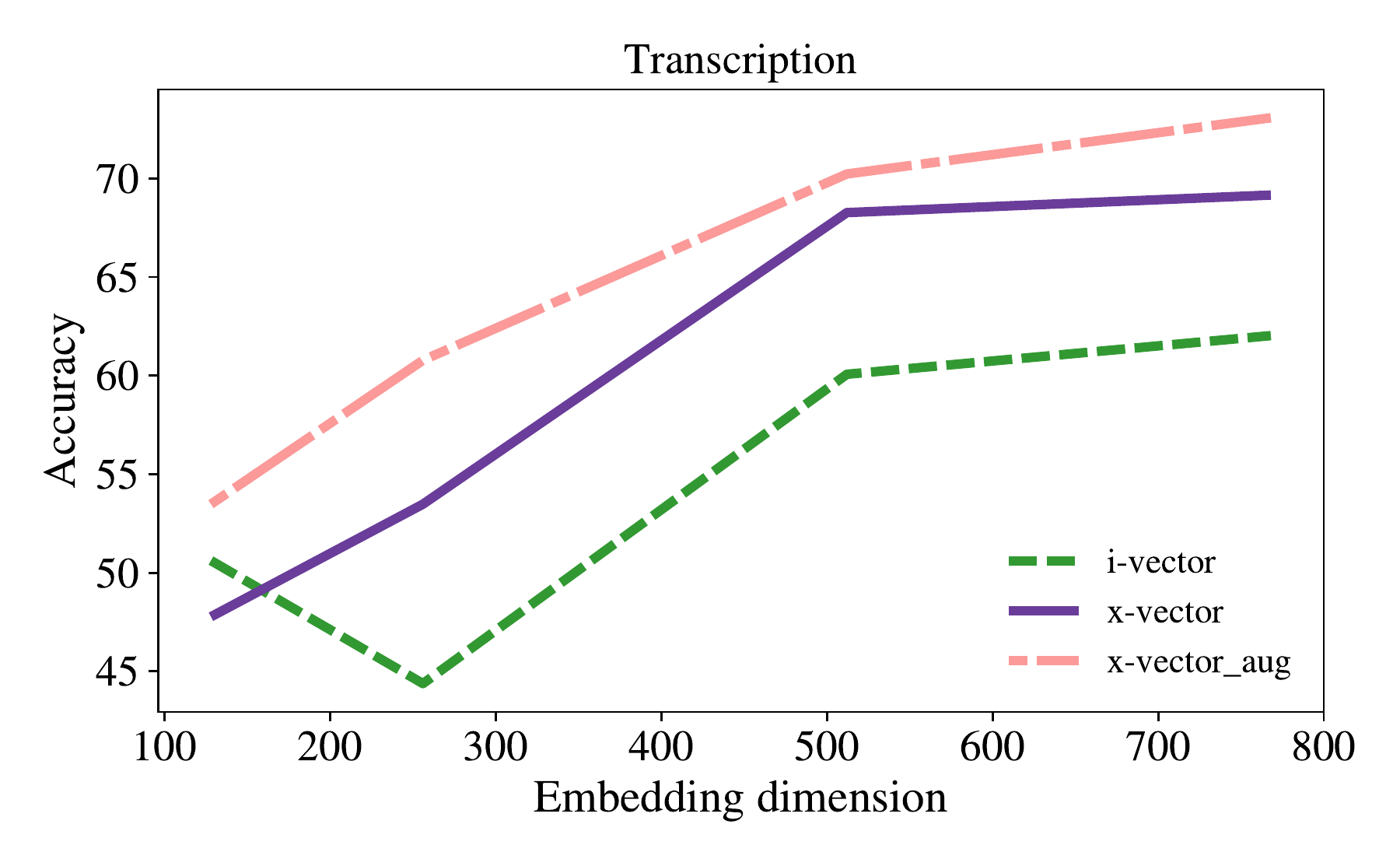}
\caption{}
\label{fig:text}
\end{subfigure}
\begin{subfigure}{0.49\linewidth}
\centering
\includegraphics[width=\linewidth]{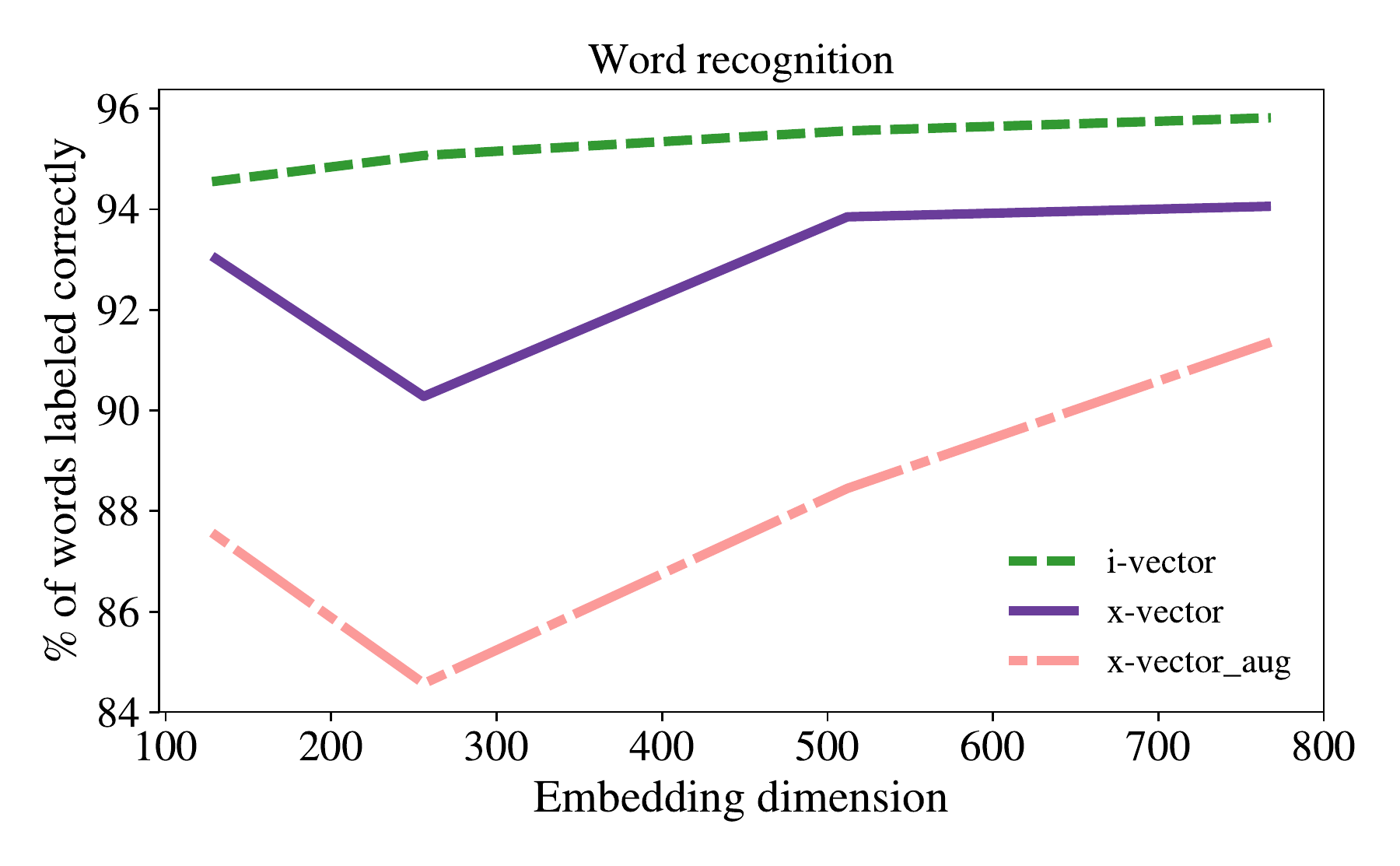}
\caption{}
\label{fig:word}
\end{subfigure}\hfill
\caption{Results for probing tasks related to text content of the utterance: (a) transcription, and (b) word recognition.}
\label{fig:probe_res_text}
\end{figure}

\begin{table}[t]
\centering
\begin{tabular}{lllllll}
\toprule
\multicolumn{1}{c}{\multirow{2}{*}{\textbf{Dim}}} & \multicolumn{3}{c}{\textbf{i-vector}} & \multicolumn{3}{c}{\textbf{x-vector}}  \\
\multicolumn{1}{c}{} & \multicolumn{1}{c}{\textbf{P}} & \multicolumn{1}{c}{\textbf{R}} & \multicolumn{1}{c}{\textbf{F}} & \multicolumn{1}{c}{\textbf{P}} & \multicolumn{1}{c}{\textbf{R}} & \multicolumn{1}{c}{\textbf{F}} \\
\midrule
\textbf{128} & 0.54 & 0.61 & 0.57 & 0.43 & 0.56 & 0.48 \\
\textbf{256} & 0.38 & 0.53 & 0.44 & 0.54 & 0.67 & 0.60 \\
\textbf{512} & 0.64 & 0.74 & 0.69 & 0.77 & 0.85 & 0.80 \\
\textbf{768} & 0.66 & 0.76 & 0.70 & 0.77 & 0.86 & 0.81 \\
\bottomrule
\end{tabular}
\caption{Performance of the embeddings in predicting the 10 common phrase utterances. P, R, and F denote precision, recall, and F-score, respectively.}
\label{tab:common_text}
\end{table}

\subsection{Speaking rate}

Since we vary speaking rates by 0.5 and 1.5, this changes speaker characteristics and is therefore a type of speaker recognition task in itself. As such, it is not surprising to find that, in Figure~\ref{fig:speed}, the embeddings achieve high accuracy in predicting the speaking rate, with x-vectors (both augmented and unaugmented) reaching close to 100\% accuracy across all dimensions.

\subsection{Transcription}

Figure~\ref{fig:text} shows the accuracy for the embeddings in predicting the utterance transcription, given the speaker embedding as input. Since several of these utterances are spoken by multiple speakers, it seems very plausible that to achieve a high recognition performance, the embedding must incidentally capture some transcription information in addition to the speaker information. In Table~\ref{tab:common_text}, we additionally report performance metrics (precision, recall, and F-score) for the 10 common-phrase utterances\footnote{Interestingly, the best performance across the board was found to be for the utterance ``Ok Google''.}. Although the x-vector DNN is trained on a large amount of unconstrained speech to predict only the speaker label, we find the x-vectors (with or without augmentation) are still not wholly invariant to lexical content, and they do retain information about what was spoken.

\begin{figure}[!t]
\begin{subfigure}{\linewidth}
\centering
\includegraphics[width=\linewidth]{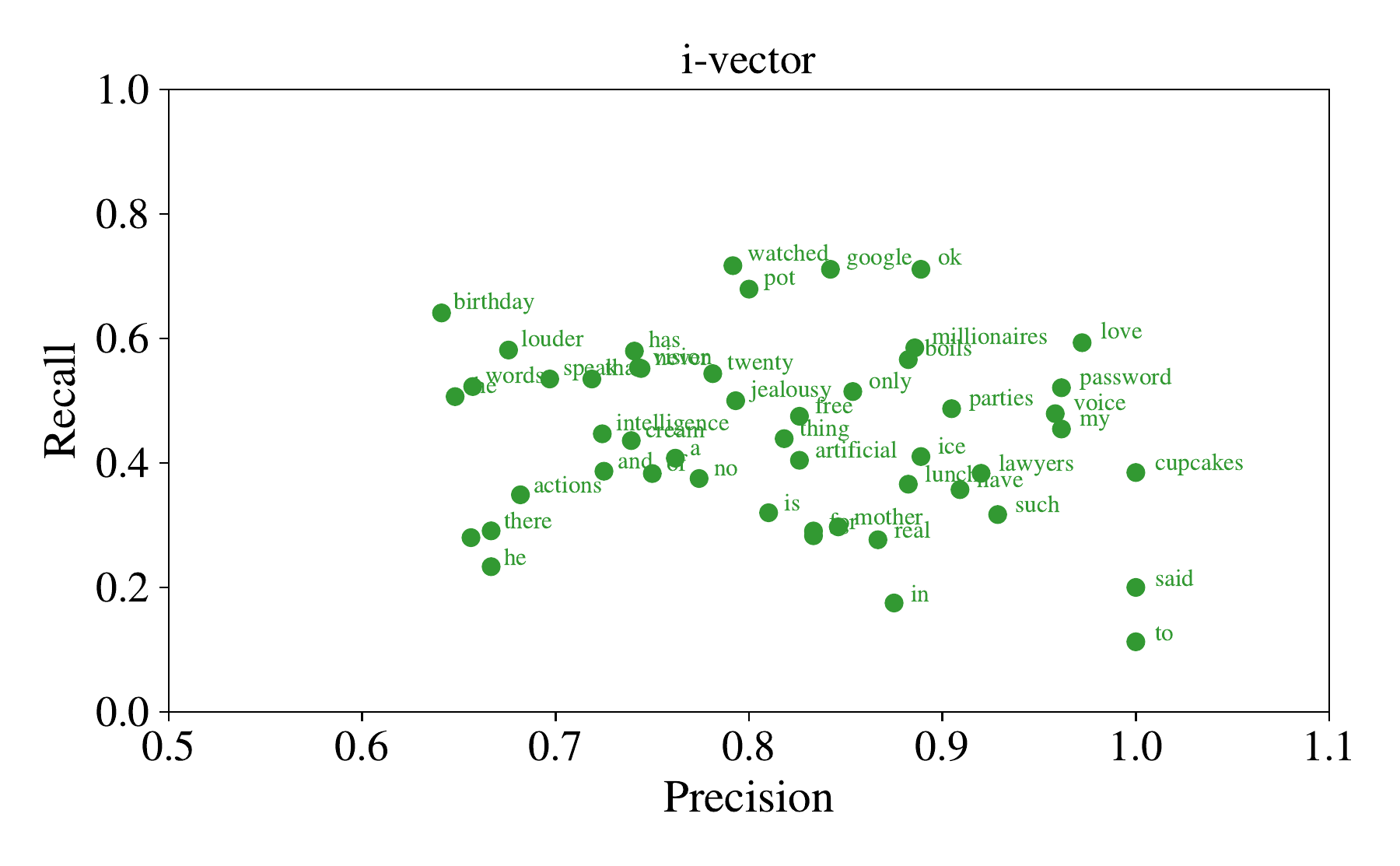}
\label{fig:ivec_word}
\end{subfigure}
\begin{subfigure}{\linewidth}
\centering
\includegraphics[width=\linewidth]{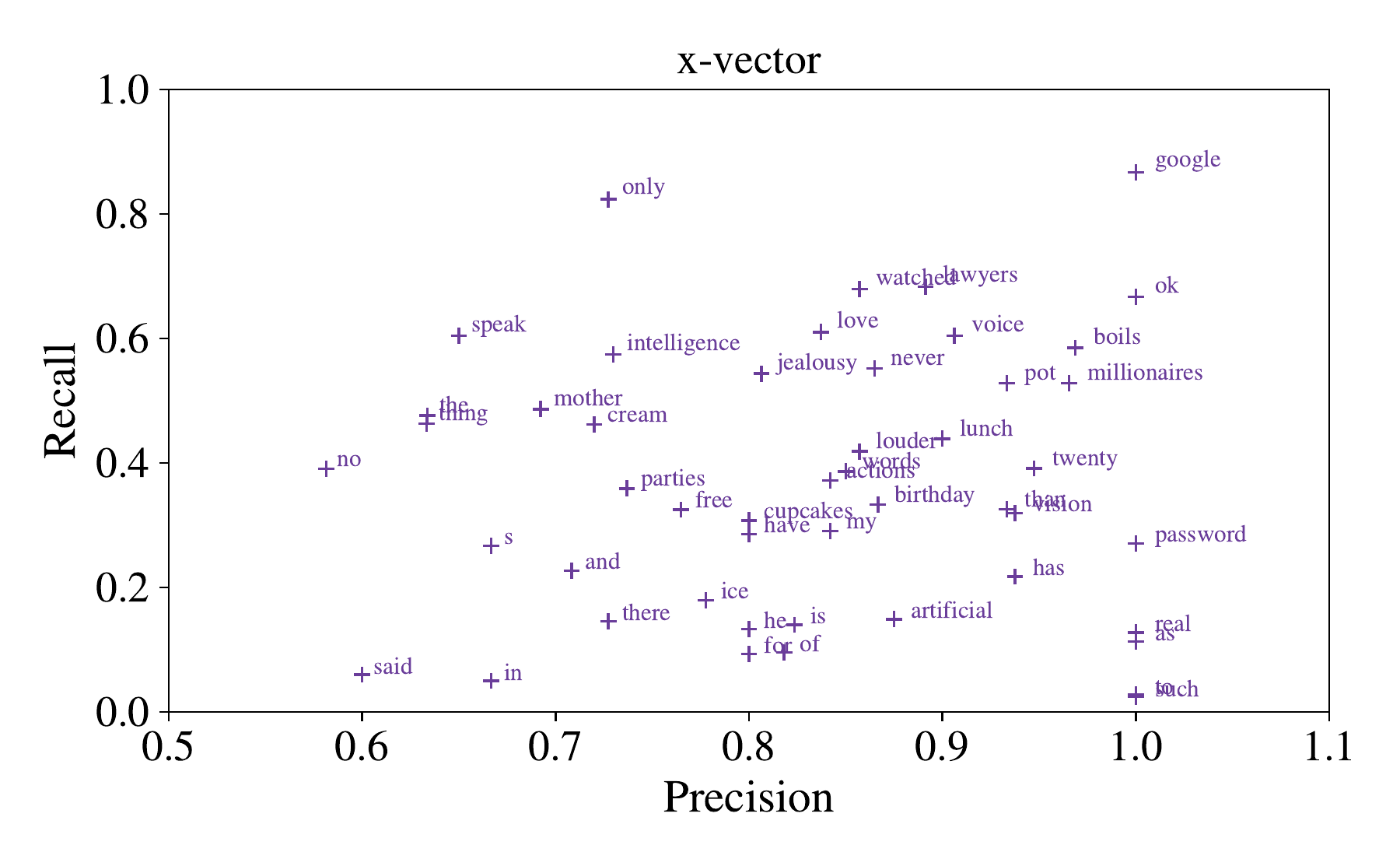}
\label{fig:xvec_word}
\end{subfigure}
\caption{Word recognition performance of speaker embeddings for top 50 most frequent words in the RedDots data set.}
\label{fig:vec_word}

\centering
\includegraphics[width=\linewidth]{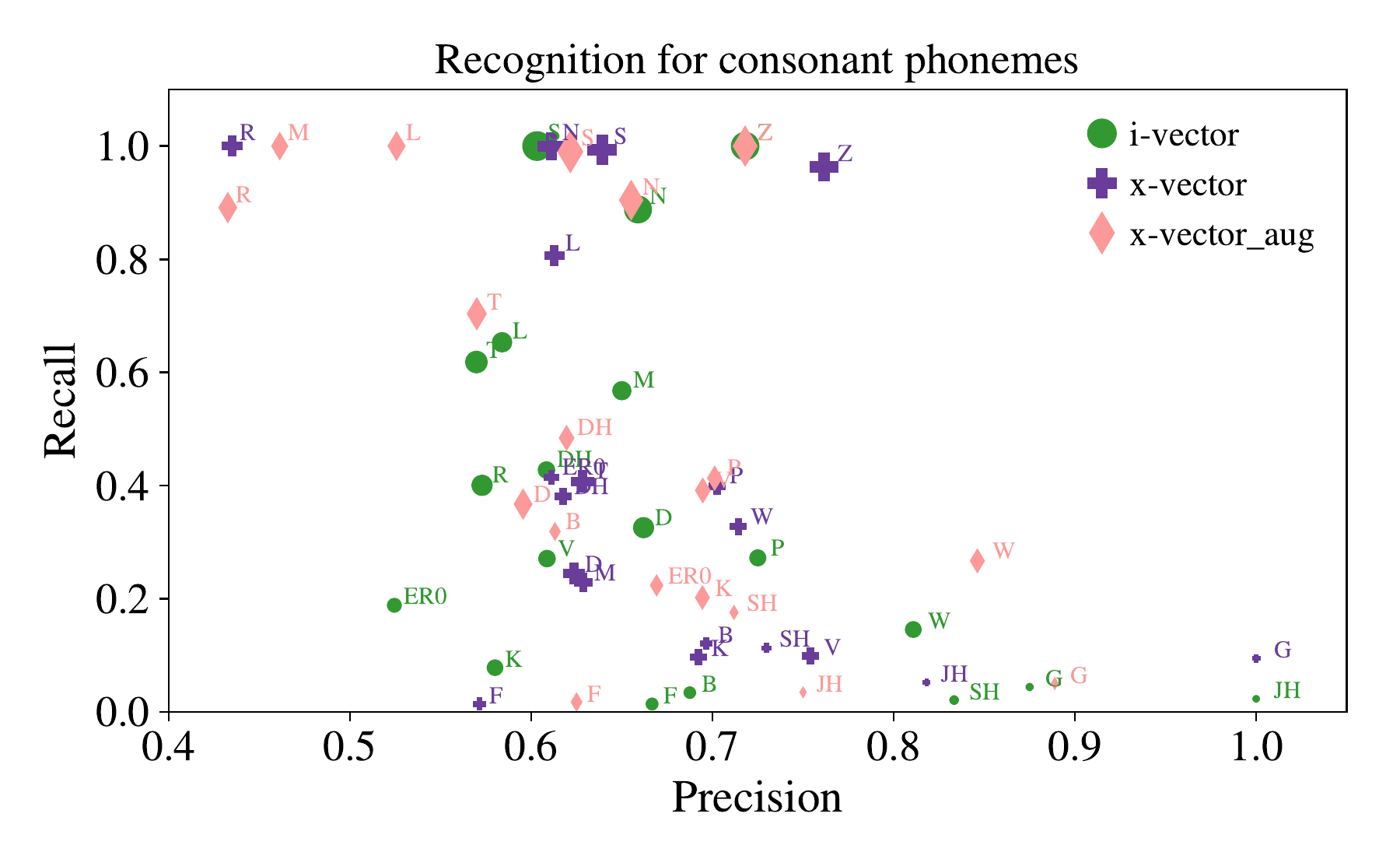}
\caption{Phoneme recognition performance for the speaker embeddings ($d = 512$) for all consonant phonemes. Marker sizes denote relative counts of the phonemes in the training set. The most recognizable phonemes for both systems are found to be \texttt{Z}, \texttt{S}, and \texttt{N}.}
\label{fig:phone}
\end{figure}

\subsection{Word recognition}

\begin{figure}[t]
\begin{subfigure}{0.49\linewidth}
\centering
\includegraphics[width=\linewidth]{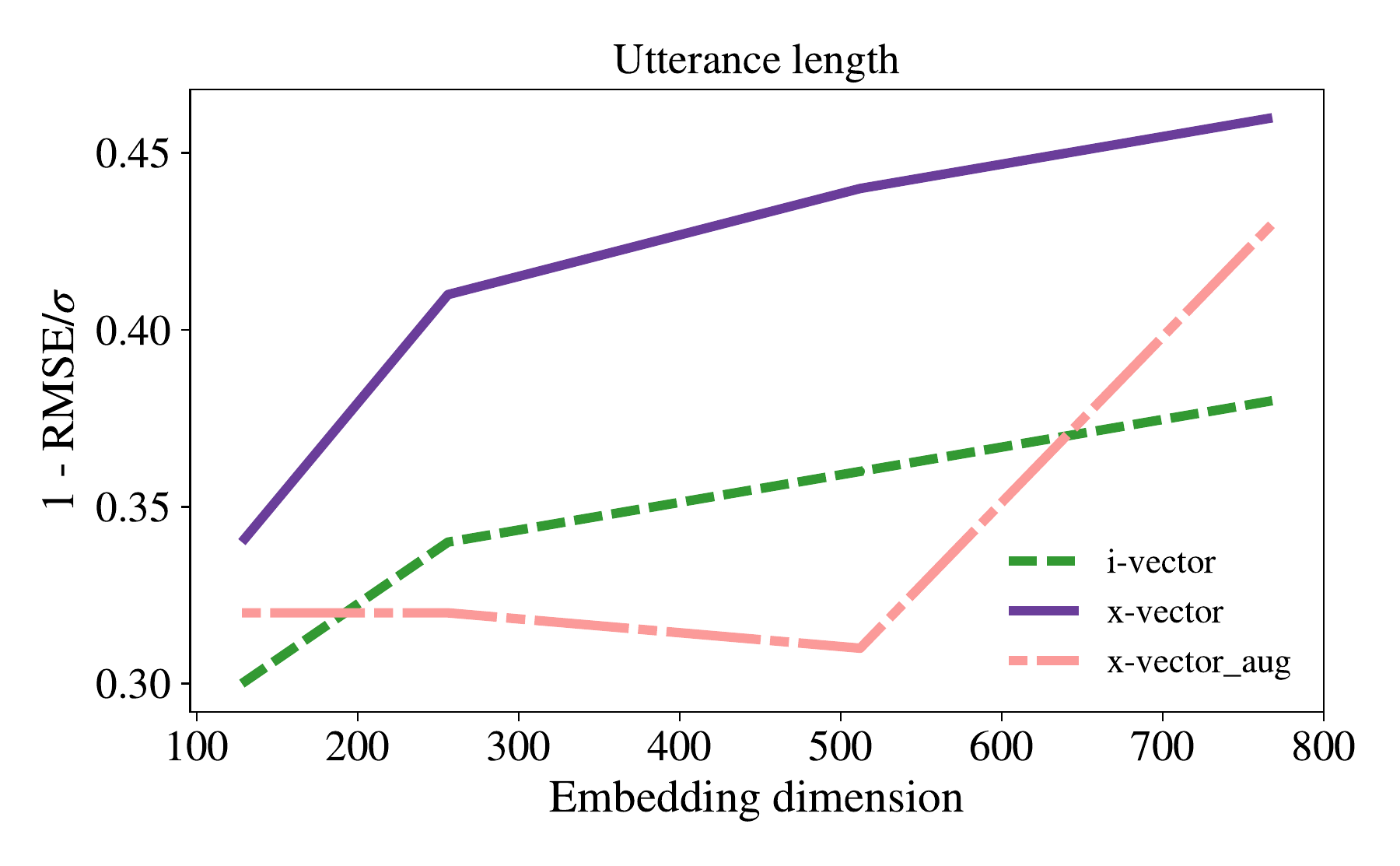}
\caption{}
\label{fig:utt_dur}
\end{subfigure}
\begin{subfigure}{0.49\linewidth}
\centering
\includegraphics[width=\linewidth]{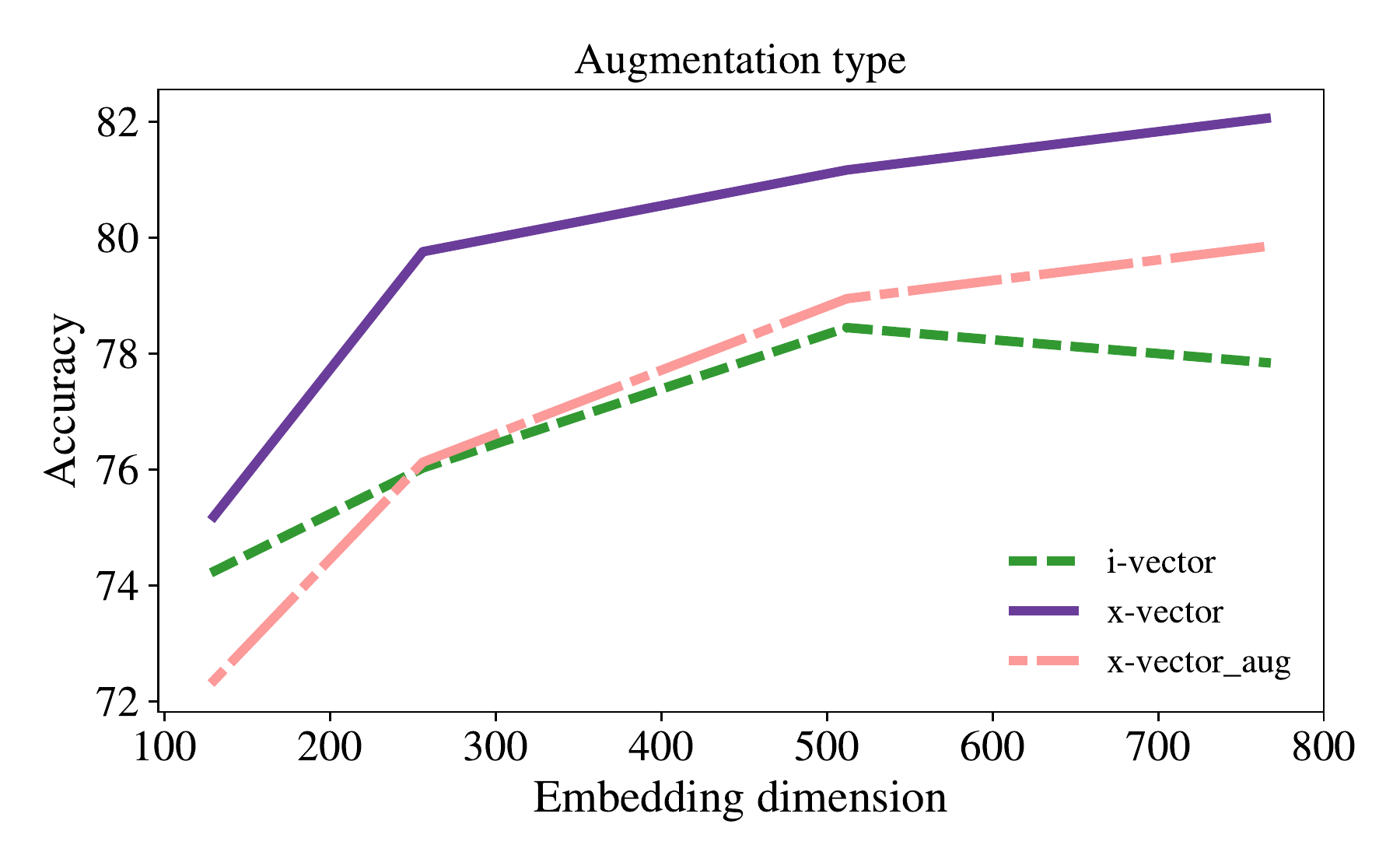}
\caption{}
\label{fig:aug_type}
\end{subfigure}\hfill
\caption{Results for probing tasks related to utterance meta information: (a) utterance length and (b) augmentation type.}
\label{fig:probe_res_meta}
\end{figure}

From Figure~\ref{fig:word}, we find that i-vectors outperform x-vectors in the word recognition task by $\sim 2$\% in terms of the average percentage of top 50 words labeled correctly across all utterances. To further investigate this result, we plot the precision and recall metrics for all 50 words in Figure~\ref{fig:vec_word}. It may be observed that x-vectors are good at recognizing words such as \textit{google}, \textit{lawyers}, \textit{password}, and so on, but fare poorly on more common words such as \textit{and}, \textit{he}, and \textit{in}. While i-vectors show a similar trend, the difference is less pronounced. Augmentation (not shown) reduces word-level recognition rate on average, but it increases recognition of keywords (most of which occurred in the common-phrase utterances), which would explain the results in Figure~\ref{fig:text} and Table~\ref{tab:common_text}.

\subsection{Phoneme recognition}

We found that the same metric as used for word recognition was not very informative in this case, since all the embeddings correctly classified around 33-34\% of the 24 consonant phonemes. Instead, we plot the precision and recall metrics for each phoneme in Figure~\ref{fig:phone}. As expected, phonemes with higher counts have better recognition rates. Nasals and plosives show similar trends across both i-vectors and x-vectors, but recognition of fricatives and approximants show high variability.



\begin{figure*}[!t]
\begin{subfigure}{0.33\linewidth}
\centering
\includegraphics[width=\linewidth]{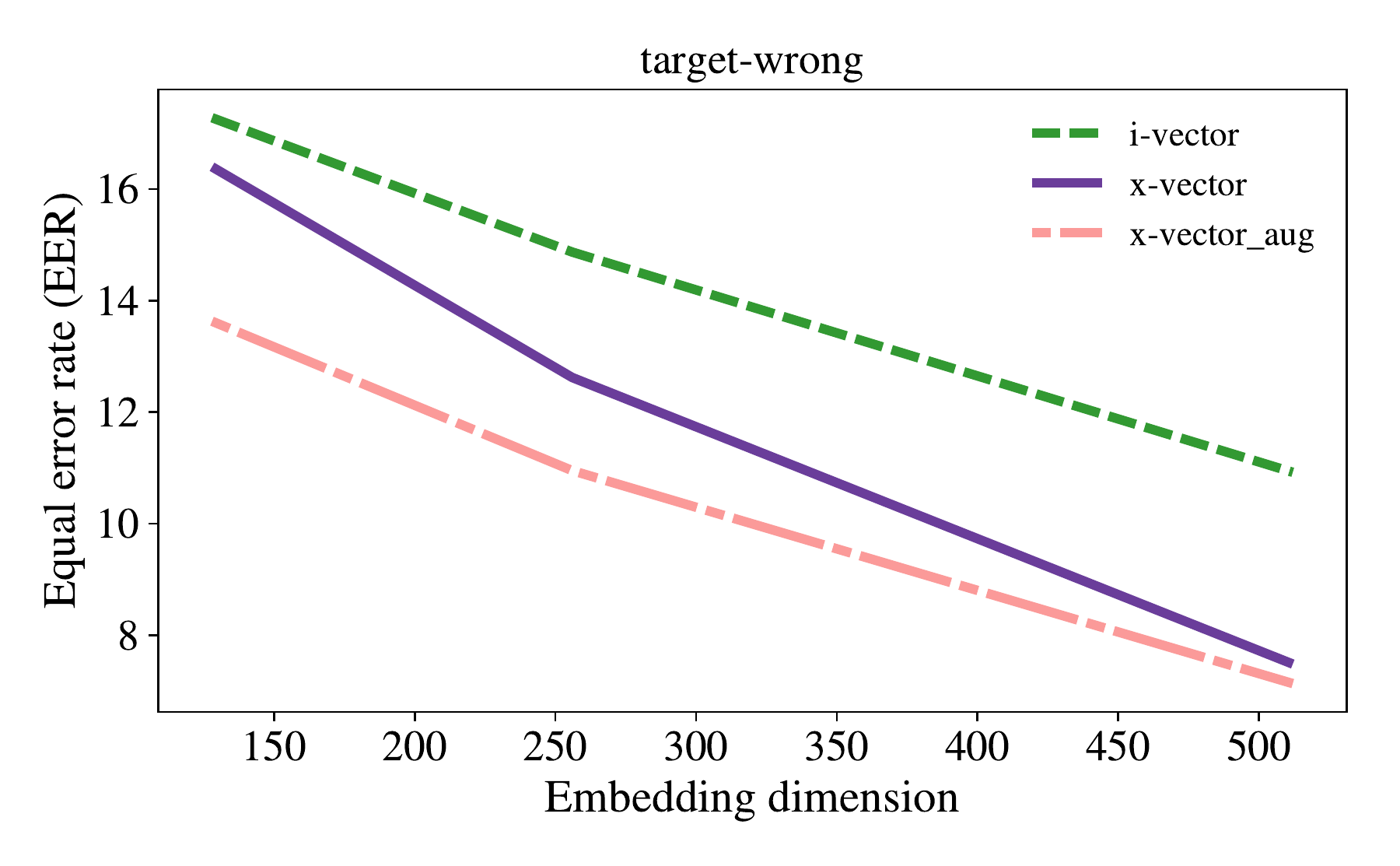}
\end{subfigure}
\begin{subfigure}{0.33\linewidth}
\centering
\includegraphics[width=\linewidth]{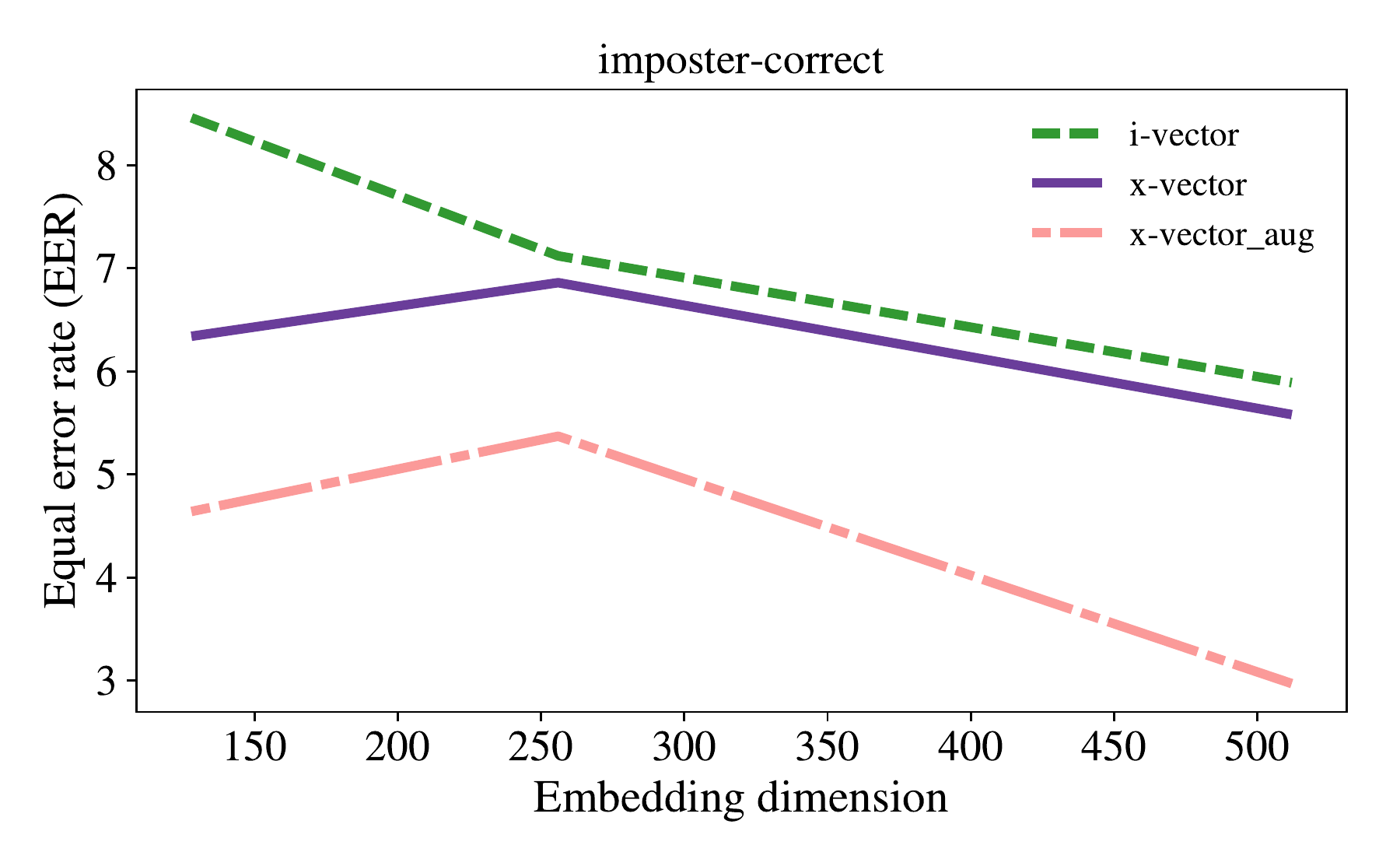}
\end{subfigure}
\begin{subfigure}{0.33\linewidth}
\centering
\includegraphics[width=\linewidth]{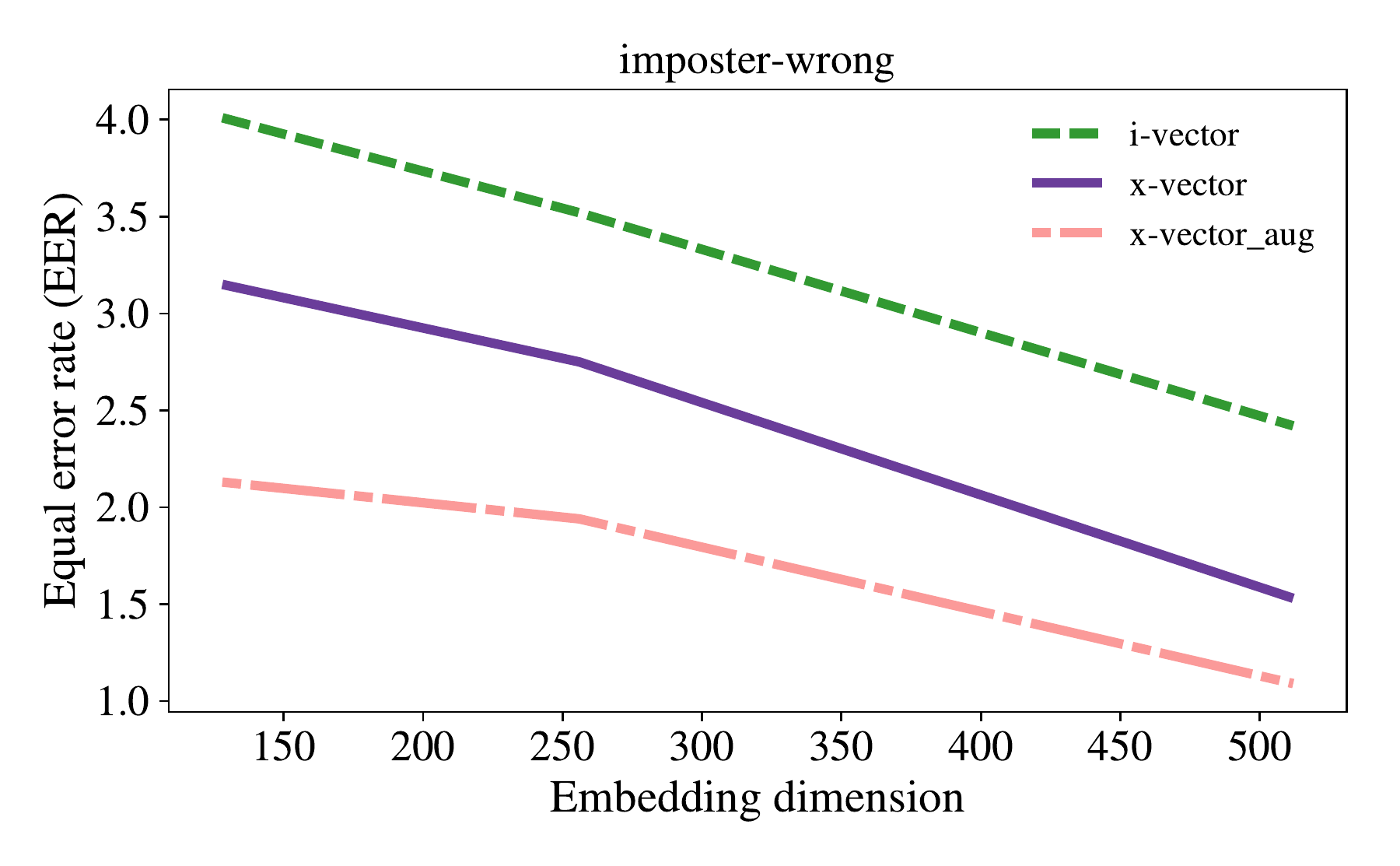}
\end{subfigure}\hfill
\caption{Performance of speaker embeddings for text-dependent speaker verification on \texttt{male\_part\_01} subset of RedDots data set.}
\label{fig:sv_task}
\end{figure*}

\subsection{Utterance length}

In this Section, we investigate whether or not information about the utterance length is retained in the speaker embeddings. From Figure~\ref{fig:utt_dur}, it can be seen that unaugmented x-vectors capture more variance in utterance length than i-vectors. We also found that for the majority of utterances, the predicted value of length is wrong by at most 0.5 seconds. Furthermore, most incorrect predictions are for very long utterances (10 or more seconds). This may be because the x-vector extractors were trained on chunks of 2 to 4 seconds, or because these utterances are outliers in the data set. Augmentation tends to decrease this quantity for x-vectors because it improves the embedding's robustness to different sources of variability.

\subsection{Augmentation type}

Figure~\ref{fig:aug_type} shows the performance of the speaker embeddings in predicting the augmentation type (i.e., clean, babble, music, or noise), and it is evident that without training augmentation, x-vectors capture more information than i-vectors in this regard. Further investigation showed that the highest recognition was for utterances with `babble', which can be attributed to the fact that adding babble creates overlapping speech which has very different acoustics than noisy speech. 

Without augmentation, x-vectors appear to capture more non-speech information than i-vectors. However, after training with augmented data, x-vectors become less sensitive to speech perturbed by noises, music or reverberation. Training with augmented data trades this noise-related information for additional speaker information. As a result of the embeddings becoming more robust, their performance on this probing task decreases.

\section{Text-dependent speaker verification}
\label{sec:reddots}

In this task, the objective is to verify a speaker based on their known utterances. In this section, we present the results for our i-vector and x-vector based systems described earlier on the RedDots data set. The data set has 62 speakers including 49 male speakers and 13 female speakers from 21 countries. The total number of sessions for the current release is 572 (473 male and 99 female sessions). The partitions I, II, and III  correspond to common, unique, and free-choice pass-phrase portions of the data set. We report results only for the \texttt{male\_part\_01} subset\footnote{All other subsets followed similar trends.}. The scores are obtained individually for the \textit{target\_wrong}, \textit{imposter\_correct}, and \textit{imposter\_wrong} cases, following the approach in \cite{zeinali2016vector}.

\subsection{Speaker verification model}

We use the speaker embedding models of dimensions 128, 256, and 512 for our experiments. In Section~\ref{sec:embeddings}, we described the methods for obtaining these embeddings. Once we have the embeddings, we use a probabilistic linear discriminant analysis (PLDA) model for the backend~\cite{prince2007probabilistic}. Similar to \cite{snyder2017deep}, we center the embeddings and reduce the dimensionality (to 100 for embeddings of dimension 128 and 256, and 200 for the 512-dimensional embeddings, respectively) using
LDA. After dimensionality reduction, the embeddings are length normalized and pairs of embeddings are compared using PLDA, which is trained on a combination of Mixer6, NIST SRE 2008, and SRE 2010 data sets~\cite{chodroff2016new, martin2009nist, martin2010nist}.

\subsection{Results}

Figure~\ref{fig:sv_task} shows the equal error rates (EERs) obtained on the three systems for the different trial types. Since all systems are trained on text-independent speaker recognition data sets, their performance on \textit{target\_wrong} trials is worse than that on the \textit{imposter\_correct} and \textit{imposter\_wrong} trials. In general, x-vectors achieve better scores than the i-vector system, and training augmentation further improves this performance by making the embeddings robust to non-speech information. Finally, increasing the dimension size improves the recognition performance across all systems.

\section{Conclusion}
\label{sec:conclusion}

We investigated the information encoded in x-vectors, compared to i-vectors, using several probing tasks. We found that in addition to speaker-related information such as session and gender, x-vectors trained on unaugmented data also capture information about the lexical content (such as keywords present in the utterance, in particular) and meta information such as utterance length and augmentation type, even in low dimensions. When augmentations are used for extractor training, some of this information is traded for better speaker recognition performance. This was confirmed by our experiments on  text-dependent speaker verification on the RedDots data set, for augmented and unaugmented x-vector systems trained in a text-independent manner. Our results motivate a line of inquiry for using x-vectors for speaker adaptation in automatic speech recognition (ASR) systems since they capture speaker and channel characteristics well.


\bibliographystyle{IEEEbib}
\bibliography{refs}

\end{document}